\documentclass[sigconf ]{acmart}

\usepackage{booktabs} 
\usepackage{multirow}
\usepackage{subfigure}
\usepackage{verbatim}
\usepackage{CJKutf8}
\usepackage{flushend}

\usepackage{makecell}
\usepackage[utf8]{inputenc}

\copyrightyear{2022}
\acmYear{2022}
\setcopyright{acmcopyright} 
\acmPrice{15.00}

\begin{document}

\title{FairRank: Fairness-aware Single-tower Ranking Framework\\ for News Recommendation}

\author{Chuhan Wu$^1$, Fangzhao Wu$^2$, Tao Qi$^1$, Yongfeng Huang$^1$}

\affiliation{%
  \institution{$^1$Department of Electronic Engineering \& BNRist, Tsinghua University, Beijing 100084 \\ $^2$Microsoft Research Asia, Beijing 100080, China}
} 
\email{{wuchuhan15,wufangzhao,taoqi.qt}@gmail.com,yfhuang@tsinghua.edu.cn}

\begin{abstract}
Single-tower models are widely used in the ranking stage of news recommendation to accurately rank candidate news according to their fine-grained relatedness with user interest indicated by user behaviors.
However, these models can easily inherit the biases related to users' sensitive attributes (e.g., gender) encoded in user behavior data for model training, and result in recommendation results unfair for users with certain attributes.
In this paper, we propose \textit{FairRank}, a fairness-aware single-tower ranking framework for news recommendation.
Since candidate news selection in existing recommender systems can be biased, we propose to use a shared candidate-aware user model to match user interest with a real displayed candidate news and a random news.
It learns a candidate-aware user embedding that reflects user interest in candidate news, and a candidate-invariant user embedding that indicates intrinsic user interest, respectively.
We apply adversarial learning to both of them to reduce the biases brought by sensitive user attributes.
In addition, we use a KL loss to regularize the attribute labels inferred from the two user embeddings to be similar, which can make the model capture less candidate-aware bias information.
Extensive experiments on two datasets show that \textit{FairRank} can improve the fairness of various single-tower news ranking models with minor performance loss.

\end{abstract}


\maketitle

\section{Introduction}

Recall and ranking are two essential steps in personalized news recommendation~\cite{wu2020mind,wu2021two}.
The recall step aims to select a small subset of candidate news  from the large news pool to comprehensively cover different user interests~\cite{qi2021hierec}.
In this step, two-tower models, which learn user interest and news embeddings independently, are common choices due to efficiency considerations~\cite{liu2020kred,ge2020graph,wu2021newsbert,wu2021personalized}.
In contrast, the ranking step needs to accurately rank candidate news according to their relevance with user interest~\cite{wang2020fine}.
Thus, single tower models that can capture the fine-grained relatedness between candidate news and user interest indicated by different user behaviors, are widely used as the ranking algorithms in news recommendation~\cite{qi2021kim,jia2021rmbert,zhang2021amm,zhang2021unbert}.

News ranking models are usually learned on users' click behavior data~\cite{lee2020news}.
In fact, click behavior data inherently encodes biases related to users' sensitive attributes such as demographics~\cite{hu2007demographic,beutel2019fairness}.
The model learned on such data can inherit and even amplify these biases, which may generate biased recommendation results that are unfair to users with certain sensitive attributes~\cite{wu2021fairness}.\footnote{We find   single-tower models may even have more severe unfairness than two-tower models, see experiments.}
Adversarial learning is a widely used technique to remove the biases related to sensitive attributes from deep models to help make fair predictions~\cite{elazar2018adversarial,zhang2018mitigating,xu2019achieving}, which also has the potential to empower fairness-aware news recommendation.
For example, \citet{wu2021fairness} proposed a decomposed adversarial learning framework that learns a bias-aware user model and a bias-free user model to capture bias information and bias-independent user interest information for news recommendation, respectively.  
However, existing methods are based on two-tower architectures that model user interest and candidate news independently, and the fairness issue of single-tower news ranking models is rarely studied.

In this paper, we propose a fairness-aware single-tower ranking framework named \textit{FairRank} for news recommendation, which can effectively improve the fairness of various single-tower news ranking models.
Since the selection of candidate news in existing recommender systems can be biased, in \textit{FairRank} we propose to use a shared candidate-aware user model to match user interest with a real displayed candidate news and a randomly selected news, respectively, to learn a candidate-aware user embedding that indicates user interest on a specific candidate news and a candidate-invariant user embedding that encodes intrinsic user interest in terms of expectation.
To reduce the bias on sensitive user attributes encoded by the model, we apply adversarial learning to both user embeddings to remove bias information.
In addition, to further eliminate bias information encoded by the selection of candidate news, we propose to use a Kullback–Leibler (KL) divergence loss to regularize the attribute labels inferred from the two user embeddings to be similar, which can encourage the model to be less sensitive to the inherent bias introduced by candidate news selection.
Extensive experiments on two real-world news recommendation datasets show that \textit{FairRank} can  improve the fairness of various single-tower ranking models with minor performance sacrifice.

\section{FairRank}\label{sec:Model}

\begin{figure}[!t]
  \centering 
      \includegraphics[width=0.99\linewidth]{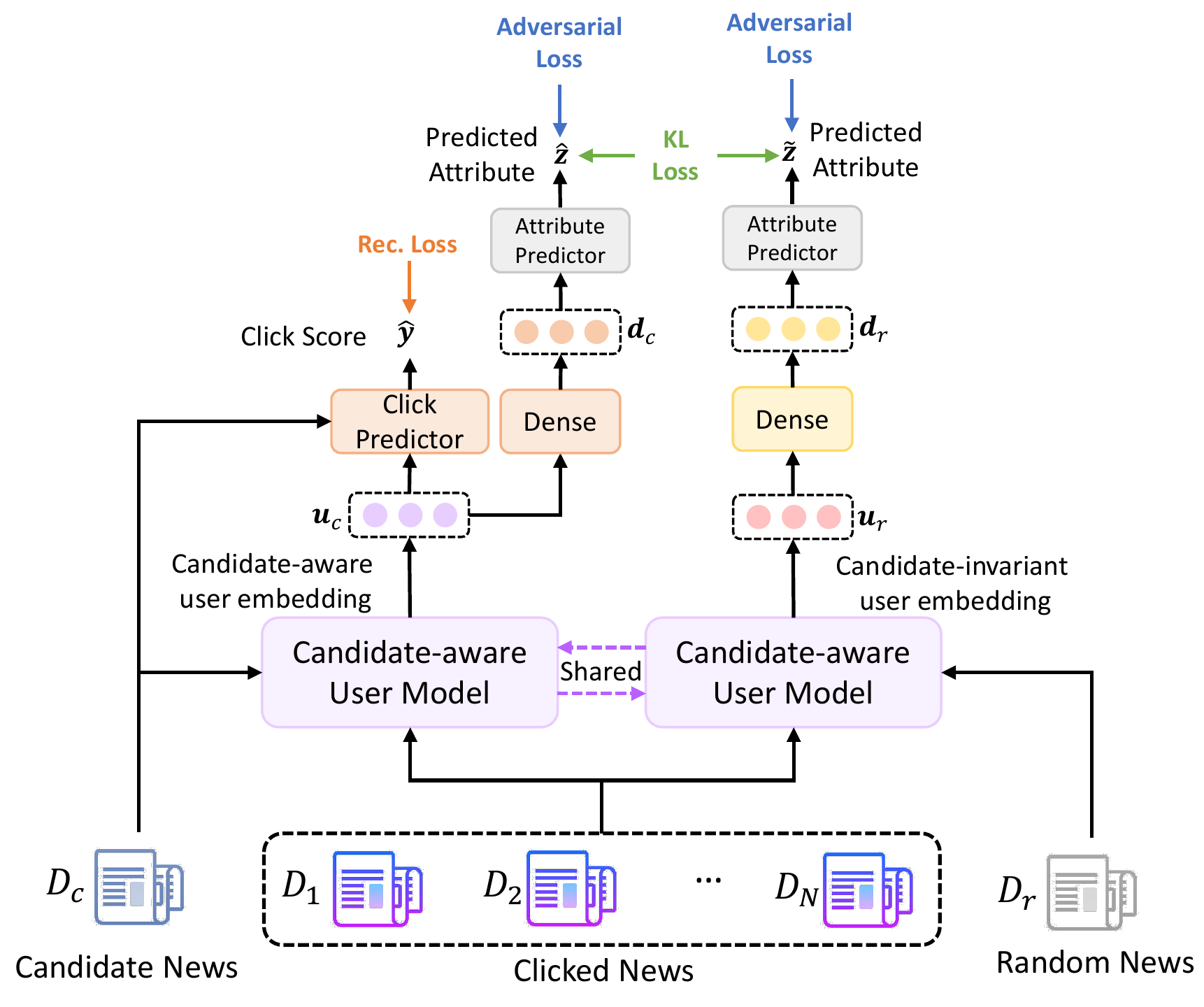} 
  \caption{The framework of \textit{FairRank}.}\label{fig.model}  
  \vspace{-0.1in}
\end{figure}

Next, we introduce the \textit{FairRank} approach for fairness-aware news ranking, as shown in  Fig.~\ref{fig.model}.
It incorporates a candidate-aware user model to capture the interactions between clicked news and a candidate news or a random news, to model user's interest in a specific candidate news and candidate-invariant user interest (in the sense of expectation). 
We then introduce its details and how to learn this model to generate fair news rankings.

\subsection{Candidate-aware Ranking Model}

Since candidate news usually can only match parts of user interest indicated by clicked news, modeling the fine-grained interactions between clicked news and candidate news may achieve better news ranking accuracy than directly matching candidate news with overall user interest.
Thus, we use a candidate-aware user model to capture users' candidate-aware interest.
It can be implemented by many architectures, such as (1) using candidate-aware attention to select clicked news according to their relevance to candidate news~\cite{wang2018dkn}; (2) using 3-D convolutional neural network to match candidate news and clicked news based on their texts~\cite{wang2020fine}; and (3) concatenating clicked news and candidate news into a long document as the input of a  Transformer~\cite{vaswani2017attention}.
It takes a user's clicked news (denoted as $[D_1, D_2, ..., D_N]$, where $N$ is the history length) and a candidate news $D_c$ as input, and outputs a candidate-aware user embedding $\mathbf{u}_c$.
This embedding is further used to compute a click score $\hat{y}$ for ranking based on the relevance between $\mathbf{u}_c$ and the hidden embedding of candidate news $D_c$ (denoted as $\mathbf{h}_c$) learned in the user model, which can be formulated as $\hat{y}=f(\mathbf{u}_c, \mathbf{h}_c)$, where $f(\cdot)$ is a relevance function that is often implemented by inner product or feedforward neural networks.

\subsection{Fairness-aware Single-tower Ranking}

We then introduce how to learn a fairness-aware single tower ranking model.
A straightforward way is applying adversarial learning to the candidate-aware user embedding to reduce biases on sensitive attributes.
Unfortunately, the candidate news list to be ranked may also be biased, and it is difficult to purify the candidate-aware user embedding.
To address this challenge, we propose to use a candidate-aware user model to jointly match clicked news with a candidate news $D_c$  and a randomly selected news $D_r$, to learn a candidate-aware user embedding $\mathbf{u}_c$ and a candidate-invariant user embedding $\mathbf{u}_r$, respectively.
Since the random news is uniformly selected from the entire news set, the candidate-invariant user embedding can reflect the intrinsic user interest in the sense of expectation.
We further apply adversarial learning to both kinds of user embeddings to learn debiased user interest.
We use two independent dense layers to learn hidden representations of them (denoted as $\mathbf{d}_c$ and $\mathbf{d}_r$) for inferring the sensitive user attribute, and a shared attribute predictor is used as the discriminator to predict sensitive attribute labels (denoted as $\mathbf{\hat{z}}$ and $\mathbf{\tilde{z}}$) as follows:
\begin{equation}
    \mathbf{\hat{z}}=\rm{softmax}(\mathbf{W}\mathbf{d}_c+\mathbf{b}),
    \end{equation}
    \begin{equation}
        \mathbf{\tilde{z}}=\rm{softmax}(\mathbf{W}\mathbf{d}_r+\mathbf{b}),
\end{equation}
where $\mathbf{W}$ and $\mathbf{b}$ are parameters in the discriminator.
We use an adversarial loss $\mathcal{L}_A$ to train the discriminator by using the crossentropy comparing $\mathbf{\hat{z}}$ or $\mathbf{\tilde{z}}$ against the attribute label.
The adversarial gradients of $\mathcal{L}_A$ will be propagated to the user model to remove its encoded biases on sensitive attributes.

However, existing studies show that adversarial learning may not fully remove sensitive attribute biases especially when the selection of  candidate news has already been biased~\cite{wu2021fairness}.
We notice that if bias information does not exist in the user model, the class distribution of the predicted attribute labels $\mathbf{\hat{z}}$ and $\mathbf{\tilde{z}}$ should be both uniform (we assume classes are balanced), which means that  $\mathbf{\hat{z}}$ and $\mathbf{\tilde{z}}$  should have the same distributions.
Thus, we apply an additional KL divergence loss $\mathcal{L}_D$ to $\mathbf{\hat{z}}$ and $\mathbf{\tilde{z}}$ to regularize them to be similar, which is formulated as follows:
\begin{equation}
    \mathcal{L}_D=KL(\mathbf{\hat{z}},\mathbf{\tilde{z}}).
\end{equation}
By optimizing this loss function, the bias distribution encoded in candidate selection is encouraged to be similar to uniform distribution, which means that less bias information can be encoded by the candidate-aware user model.

\subsection{Model Training}

We finally introduce the learning of \textit{FairRank}.
For the main news ranking task, following~\cite{wu2019npa} we use negative sampling strategies to build training samples and use the InfoNCE~\cite{oord2018representation} loss for optimization.
The unified loss $\mathcal{L}$ for model training is  as follows:
\begin{equation}
    \mathcal{L}=\mathcal{L}_R-\lambda \mathcal{L}_A+ \mathcal{L}_D,
\end{equation}
where $\lambda$ is a coefficient to control the relative importance of adversarial gradients.

\section{Experiments}\label{sec:Experiments}

\subsection{Datasets and Experimental Settings}
We use two real-world datasets for experiments.
The first one is the dataset used in~\cite{wu2021fairness} (denoted as \textit{FairNews}).
There are 2,484 male users and 1,744 female
users with observed gender labels.
The second dataset is constructed by ourselves by collecting users' click logs between 06/23/2019 and 07/20/2019 on a commercial news feeds platform (denoted as \textit{FairFeeds}).
There are 32,124 male users and 17,876 female users with gender labels.
The logs in the first 3 weeks are for training and validation (9:1 split), and the rest are for test. 
The statistics of both datasets are listed in Table~\ref{table.dataset}.

\begin{table}[h]
\centering

	\caption{Statistics of two news datasets.}\label{table.dataset}
\resizebox{0.48\textwidth}{!}{
\begin{tabular}{cccccc}
\Xhline{1.5pt}
               & \textbf{\#News} & \textbf{\#Users} & \textbf{\#Impressions} & \textbf{\#Clicks}  \\ \hline
\textit{FairNews}  & 42,255  & 10,000  & 360,428    &  503,698              \\
\textit{FairFeeds} & 112,052 & 337,286   & 500,000        & 853,291               \\ 
\Xhline{1.5pt}
\end{tabular}
}
\end{table}

In our experiments we use Adam~\cite{kingma2014adam} as the optimizer, and the learning rate is 1e-4. 
The adversarial loss weight $\lambda$ is 0.5.
The batch size is 32.
We search hyperparameters according to validation results.
Motivated by~\cite{wu2021fairness}, we choose AUC and nDCG@10 to evaluate ranking accuracy on impression data, and use the accuracy  of gender prediction from the top 10 and 20 ranked news within 100 randomly selected candidate news as fairness metrics.
These metrics are indications of casual fairness~\cite{zhang2018fairness}.
We report the average results of 5 independent experiments.

\subsection{Fairness Comparison between Single-tower and Two-tower Models}

\begin{table}[t]

\caption{Fairness comparison between different single-tower and two-tower models.}\label{st}
\resizebox{0.95\linewidth}{!}{
\begin{tabular}{lcccc}
\Xhline{1pt}
\multicolumn{1}{c}{\multirow{2}{*}{\textbf{Methods}}} & \multicolumn{2}{c}{\textbf{FairNews}} & \multicolumn{2}{c}{\textbf{FairFeeds}} \\
\multicolumn{1}{c}{}                                  & Acc@10            & Acc@20            & Acc@10             & Acc@20            \\ \hline
DKN                                                   & 56.72             & 57.87             & 57.69              & 58.53             \\
DAN                                                   & 57.50             & 58.38             & 58.20              & 59.11             \\
FIM                                                   & 58.32             & 59.64             & 59.33              & 60.17             \\
UNBERT                                                & 58.98             & 60.16             & 59.92              & 60.84             \\ \hline
EBNR                                                  & 54.85             & 56.10             & 55.24              & 56.33             \\
NAML                                                  & 56.22             & 56.83             & 56.84              & 57.40             \\
NPA                                                   & 56.42             & 57.01             & 57.11              & 57.82             \\
NRMS                                                  & 56.68             & 57.54             & 57.22              & 57.96             \\
PLM-NR                                                & 56.98             & 57.84             & 57.59              & 58.20             \\ \Xhline{1pt}
\end{tabular}
}
\end{table}

We first compare the recommendation fairness of different single-tower and two-tower methods.
The compared single-tower models include: 
(1) \textit{DKN}~\cite{wang2018dkn}, a news recommendation method based on candidate-aware attention; 
(2) \textit{DAN}~\cite{zhu2019dan}, user modeling candidate-aware attention and attentional LSTM; 
(3) \textit{FIM}~\cite{wang2020fine}, a fine-grained interest matching approach for news recommendation; 
(4) \textit{UNBERT}~\cite{zhang2021unbert}, a BERT-based news recommendation method that concatenates candidate news and clicked news into a single document.
The compared two-tower models include:
(1) \textit{EBNR}~\cite{okura2017embedding}, embedding-based news recommendation with GRU networks;
(2)\textit{NAML}~\cite{wu2019}, attentive multi-view learning for news recommendation;
(3) \textit{NPA}~\cite{wu2019npa}, news recommendation with personalized attention;
(4) \textit{NRMS}~\cite{wu2019nrms}, multi-head self-attention networks for news recommendation; 
(5) \textit{PLM-NR}~\cite{wu2021empowering}, pretrained language model empowered news modeling for news recommendation.
The results are shown in Table~\ref{st}.
We find that most single-tower ranking models generate unfairer recommendation results than two-tower models.
This may be because single-tower models conduct finer-grained matching between user behaviors and the candidate~\cite{yu2021cross}, and thereby has a higher risks of finding shortcuts and encoding bias information.
It shows that single-tower ranking models are suffered more from the biases encoded in user behavior data, and learning fair single-tower ranking models is more challenging.

\begin{table}[t]
\centering

\caption{Ranking fairness of different methods. Lower scores indicate better fairness.}  \label{p2}
\resizebox{0.95\linewidth}{!}{
\begin{tabular}{lcccc}
\Xhline{1pt}
\multicolumn{1}{c}{\multirow{2}{*}{\textbf{Methods}}} & \multicolumn{2}{c}{\textbf{FairNews}} & \multicolumn{2}{c}{\textbf{FairFeeds}} \\
\multicolumn{1}{c}{}                                  & Acc@10            & Acc@20            & Acc@10             & Acc@20            \\ \hline
DKN                                                   & 56.72             & 57.87             & 57.69              & 58.53             \\
DKN+AL                                                & 55.46             & 56.50             & 56.45              & 57.20             \\
DKN+FairRec                                           & 54.95             & 56.12             & 56.13              & 56.88             \\
DKN+FairRank                                          & 52.42             & 53.50             & 52.92              & 53.70             \\ \hline
DAN                                                   & 57.50             & 58.38             & 58.20              & 59.11             \\
DAN+AL                                                & 55.92             & 57.09             & 56.46              & 57.32             \\
DAN+FairRec                                           & 55.00             & 55.39             & 55.81              & 56.43             \\
DAN+FairRank                                          & 52.62             & 53.66             & 52.95              & 53.68             \\ \hline
FIM                                                   & 58.32             & 59.64             & 59.33              & 60.17             \\
FIM+AL                                                & 56.50             & 57.72             & 57.25              & 57.98             \\
FIM+FairRec                                           & 55.65             & 56.33             & 55.94              & 56.69             \\
FIM+FairRank                                          & 53.06             & 53.78             & 53.29              & 53.88             \\ \hline
UNBERT                                                & 58.98             & 60.16             & 59.92              & 60.84             \\
UNBERT+AL                                             & 57.47             & 58.81             & 57.92              & 58.80             \\
UNBERT+FairRec                                        & 56.21             & 56.64             & 56.87              & 57.44             \\
UNBERT+FairRank                                       & 53.70             & 54.24             & 53.96              & 54.55             \\ \Xhline{1pt}
\end{tabular}
}
\label{result}
\end{table}

 \begin{table}[t]
\centering
\caption{Ranking accuracy of different methods. Higher scores indicate better accuracy.}  \label{p1}
\resizebox{0.95\linewidth}{!}{
\begin{tabular}{lcccc}
\Xhline{1pt}
\multicolumn{1}{c}{\multirow{2}{*}{\textbf{Methods}}} & \multicolumn{2}{c}{\textbf{FairNews}} & \multicolumn{2}{c}{\textbf{FairFeeds}} \\ 
\multicolumn{1}{c}{}                                  & AUC               & nDCG@10            & AUC               & nDCG@10            \\ \hline
DKN                                                    & 60.36             & 38.83              & 65.24             & 32.10              \\
DKN+AL                                                 & 59.97             & 38.48              & 64.76             & 31.88              \\
DKN+FairRec                                            & 59.83             & 38.39              & 64.65             & 31.82              \\
DKN+FairRank                                           & 59.91             & 38.44              & 64.69             & 31.86              \\ \hline
DAN                                                    & 61.40             & 39.78              & 66.12             & 32.56              \\
DAN+AL                                                 & 60.82             & 39.25              & 65.59             & 32.29              \\
DAN+FairRec                                            & 60.84             & 39.26              & 65.48             & 32.23              \\
DAN+FairRank                                           & 60.92             & 39.31              & 65.66             & 32.34              \\ \hline
FIM                                                    & 63.23             & 41.48              & 67.15             & 33.18              \\
FIM+AL                                                 & 62.57             & 40.93              & 66.52             & 32.84              \\
FIM+FairRec                                            & 62.64             & 40.98              & 66.60             & 32.88              \\
FIM+FairRank                                           & 62.73             & 41.06              & 66.70             & 32.94              \\ \hline
UNBERT                                                 & 64.40             & 42.45              & 68.33             & 33.96              \\
UNBERT+AL                                              & 63.87             & 41.90              & 67.67             & 33.60              \\
UNBERT+FairRec                                         & 63.81             & 41.85              & 67.70             & 33.64              \\
UNBERT+FairRank                                        & 63.96             & 41.98              & 67.82             & 33.72              \\\Xhline{1pt}
\end{tabular}
}

\label{result}
\end{table}

 \begin{figure}[!t]
	\centering 
	\subfigure[\textit{FairNews}.]{
	\includegraphics[width=0.75\linewidth]{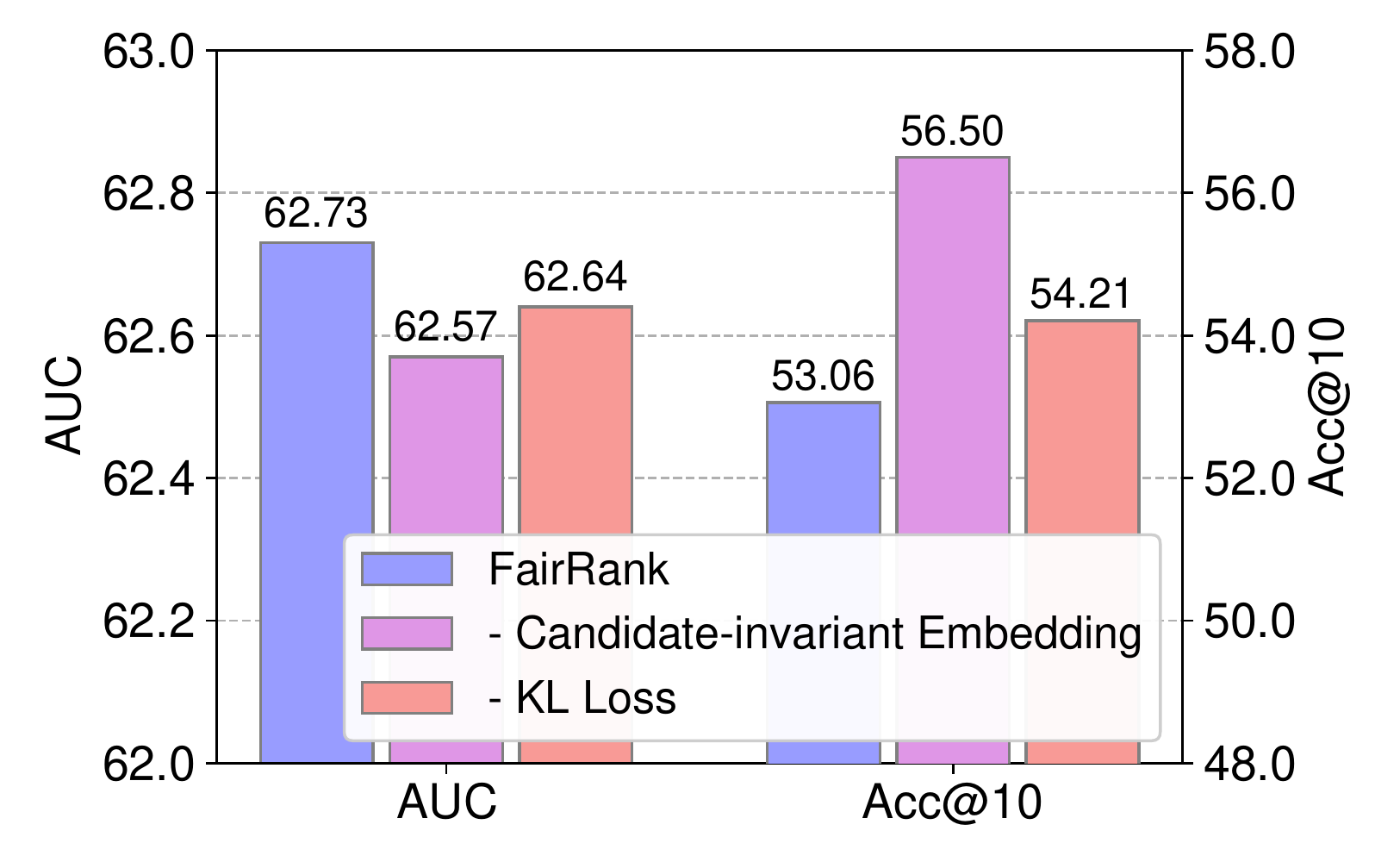} 
	}
		\subfigure[\textit{FairFeeds}.]{
		\includegraphics[width=0.75\linewidth]{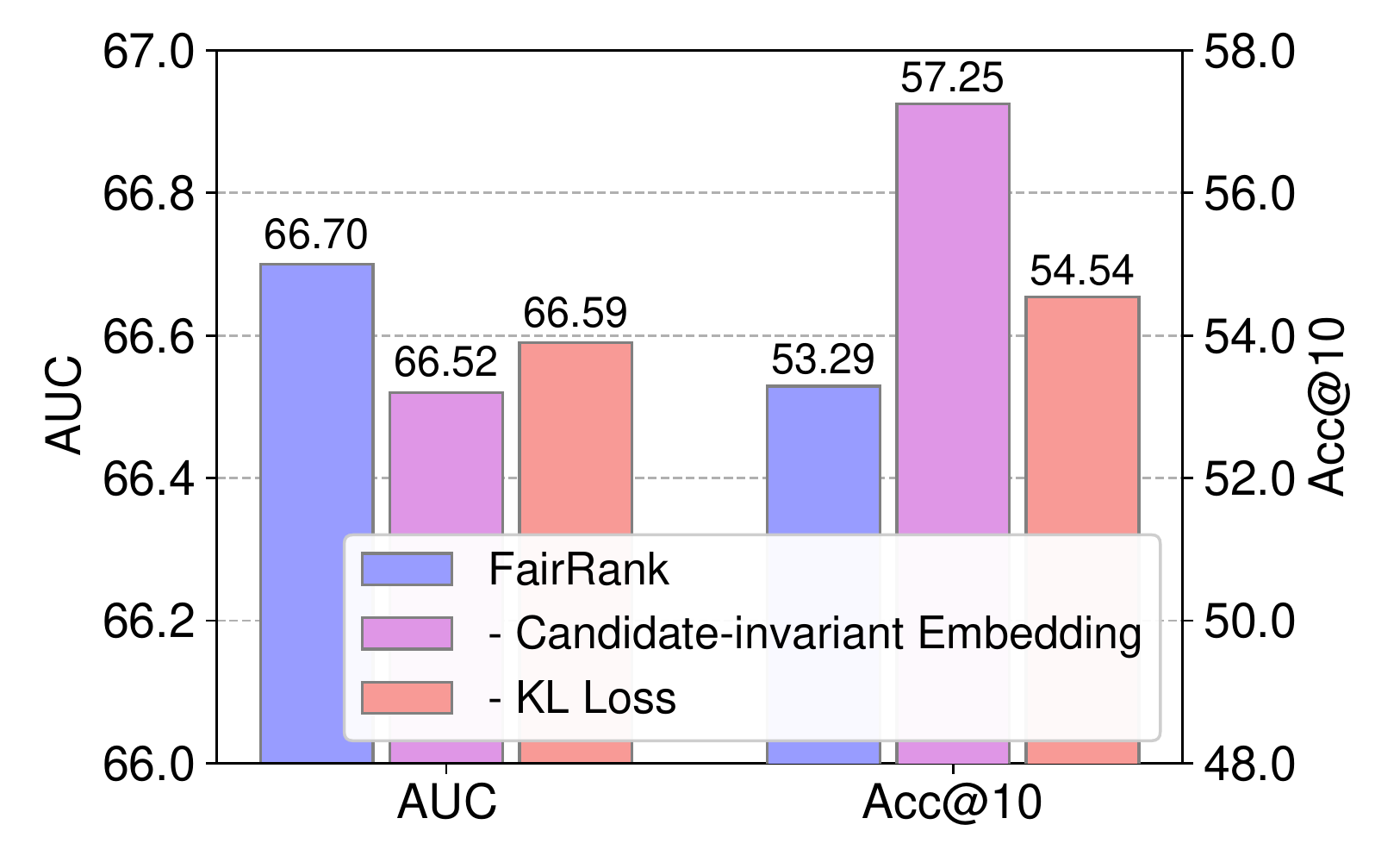} 
		}
\caption{Effectiveness of the candidate-invariant user embedding and KL loss.}\label{fig.ab1}\vspace{-0.1in}
\end{figure}
 \begin{figure}[!t]
	\centering 
		\subfigure[\textit{FairNews}.]{
	\includegraphics[width=0.75\linewidth]{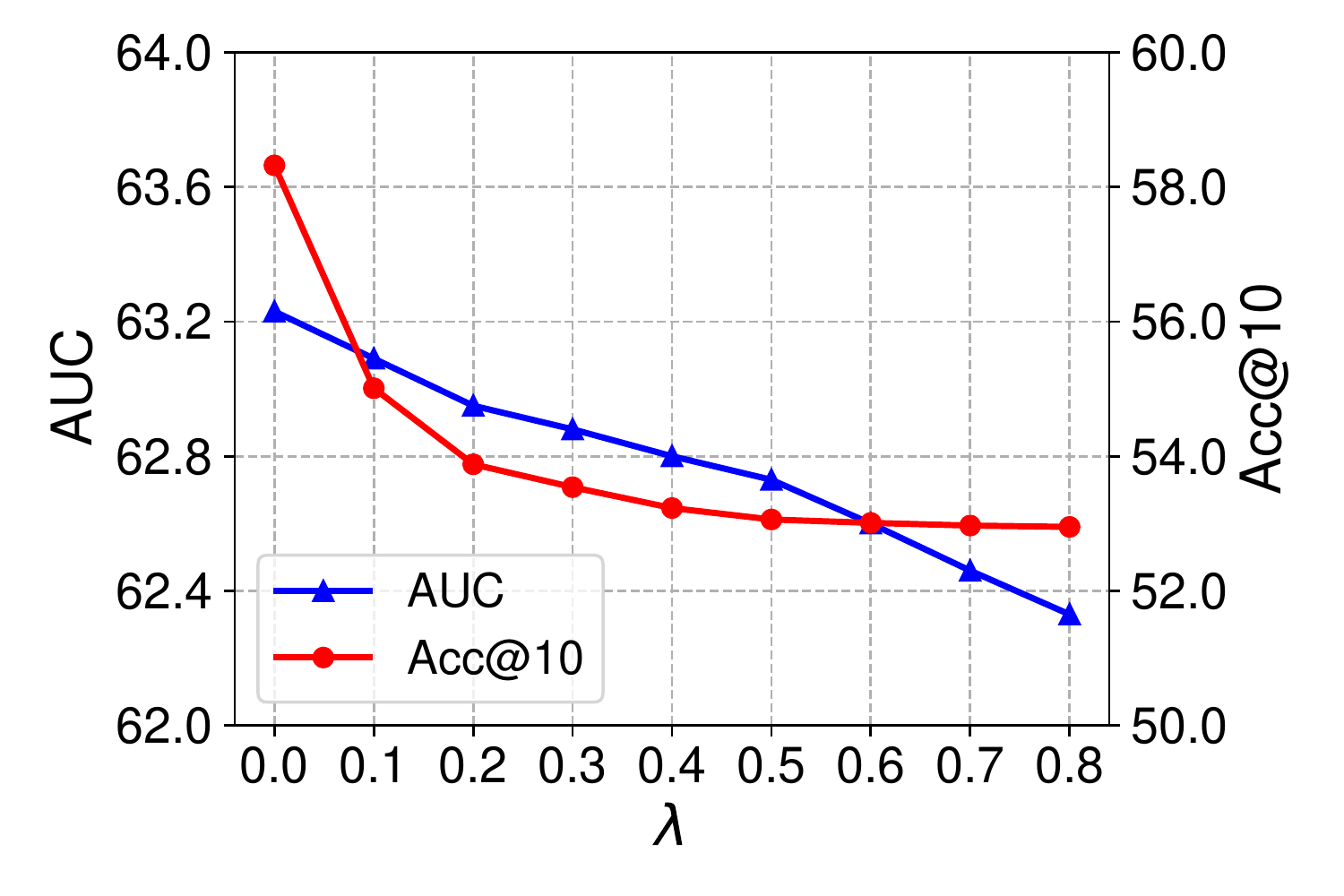} 
	}
		\subfigure[\textit{FairFeeds}.]{
		\includegraphics[width=0.75\linewidth]{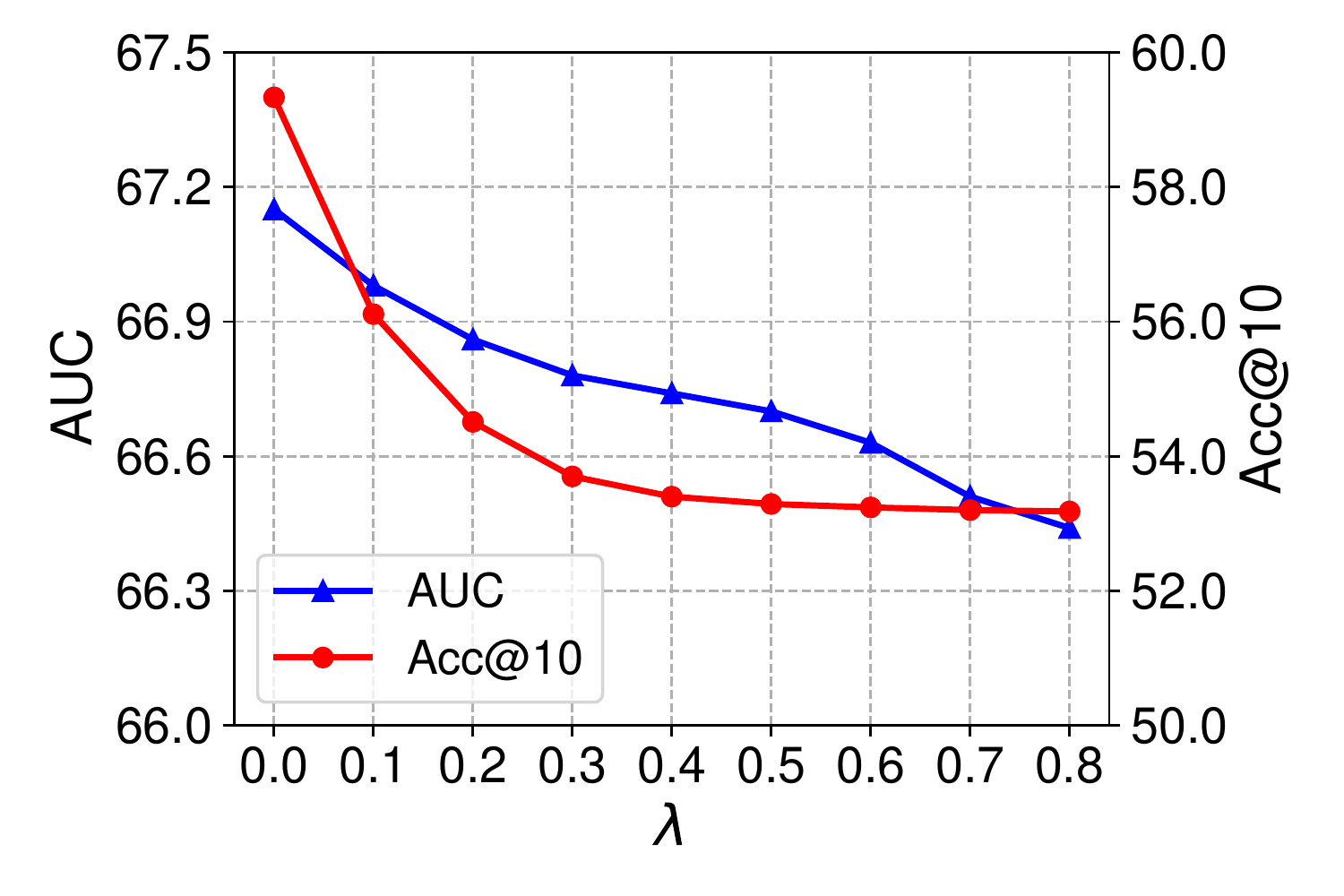} 
		}
\caption{Influence of the coefficient $\lambda$.}\label{fig.d1}
\end{figure}

\subsection{Performance Comparison}

Next, we verify whether \textit{FairRank} can effectively mitigate the heavy unfairness issue of single-tower news ranking framework.
We compare \textit{FairRank} with two fairness-aware methods, i.e., (1) \textit{AL}~\cite{beutel2017data}, applying adversarial learning to the candidate-aware user embedding; (2) \textit{FairRec}~\cite{wu2021fairness}, applying the decomposed adversarial learning method to the candidate-aware user model.
The ranking fairness and accuracy of different methods based on different basic models are shown in Tables~\ref{p2} and~\ref{p1}, respectively.
From the results, we find all basic single-tower models are heavily affected by the biases related to sensitive attributes, and their ranking results are unfair.
In addition, both \textit{AL} and \textit{FairRec} cannot effectively reduce the biases encoded in ranking results.
This is because the selection of candidate news within an impression can be biased, and it is difficult for the single-tower models to remove their biases.
By contrast, our \textit{FairRank} approach can achieve much better fairness and even slightly better accuracy than \textit{AL} and \textit{FairRec}, showing that \textit{FairRank} is more appropriate for learning fair news ranking models.

\subsection{Ablation Study}

We then study the influence of applying adversarial learning to the candidate-invariant user embedding and the additional KL regularization loss by removing them from \textit{FairRank} to compare the accuracy and fairness changes.
We use \textit{FIM} as the basic model, and the results on \textit{FairNews} are shown in Fig.~\ref{fig.ab1}.
We find that removing bias information from the candidate-invariant user embedding is important for improving fairness, since it is free from the interference of biases encoded by impression data and can better remove the intrinsic biases captured by the user model.
In addition, the KL loss can also improve ranking fairness.
This may be because it can help the model to encode less bias information to make the attribute labels inferred from both candidate-aware and candidate-invariant user embeddings to be uniform.

\subsection{Hyperparameter Analysis}

Finally, we discuss the impact of adversarial loss weight $\lambda$ on accuracy and fairness.
The results on \textit{FairNews} are shown in Fig.~\ref{fig.d1}.
When $\lambda$ is larger, the accuracy usually decreases because the adversarial loss will influence the main recommendation task, while the fairness score does not have notable improvements when $\lambda$ is larger than 0.5.
This is because the main recommendation task cannot gain sufficient supervision signals when the adversarial loss is too strong. 
Thus, we choose $\lambda=0.5$ to achieve good fairness without hurting too much ranking accuracy.
\section{Conclusion}\label{sec:Conclusion}

In this paper, we propose a fairness-aware single-tower ranking framework for news recommendation, named \textit{FairRank}.
We propose to use a candidate-aware user model to match clicked news with both candidate news and random news to learn a candidate-aware user embedding and a candidate-invariant user embedding, respectively, to help reduce the bias introduced by candidate news selection.
We apply adversarial learning to them to remove bias information, and further add a KL loss that regularizes the attribute labels inferred from the two user embeddings to be similar to make the user model  less bias-sensitive.
Experiments on two datasets show that \textit{FairRank} can effectively improve the  fairness of many single-tower news ranking models with minor accuracy losses.

\bibliographystyle{ACM-Reference-Format}
\bibliography{main}

\end{document}